\documentstyle[seceq,epsf,mbf,wrapft,preprint]{ptptex}

\title{Linear $\Sigma$ Model in the Gaussian Functional Approximation}
\author{Issei {\sc Nakamura} and V. {\sc Dmitra\v sinovi\' c}
\footnote{Address after 1st July 2001:
Vin\v ca Institute of Nuclear Sciences,
P. O. Box 522, 11001 Belgrade, Yugoslavia.}}
\inst{Research Center for Nuclear Physics, Osaka 
University, \\
Ibaraki, 567-0047, Japan}
\recdate{June 15, 2001}

\abst{We apply a self-consistent relativistic mean-field 
variational ``Gaussian functional'' (or Hartree)
approximation to the linear $\sigma$ model with spontaneously
and explicitly broken chiral $O(4)$ symmetry.
We set up the self-consistency, or ``gap'' and the 
Bethe-Salpeter equations. 
We check and confirm the chiral Ward-Takahashi identities, among them
the Nambu-Goldstone theorem and the (partial) axial current conservation [CAC],
both in and away from the chiral limit. With explicit chiral symmetry breaking 
we confirm the Dashen relation for the pion mass and partial CAC. 
We solve numerically the gap and Bethe-Salpeter equations, discuss 
the solutions' properties and the particle content of the theory.}
\begin{document}

\maketitle

\section{Introduction}
\label{sec:intro}

The Gell-Mann--Levy [GML] linear sigma model has long been a 
subject of nonperturbative studies, both for its particle physics 
and statistical mechanics applications. \cite{lee72,ls90}
In this paper we apply a new chirally invariant version of the Lorentz 
invariant self-consistent mean-field variational approximation 
that goes by many names, {\it inter alia} the Gaussian functional approximation 
\cite{bg80,hat92} to the linear sigma model. 
The improvement that we present in this paper is the
correct implementation of the chiral symmetry in this approximation. 
We prove the chiral Ward-Takahashi identities, 
among them the Nambu-Goldstone theorem, the Dashen
relation, and the axial current (partial) conservation (PCAC)
in this approximation. Then we present a numerically obtained
solution of the gap and Bethe-Salpeter 
equations and discuss the particle content of the theory in this 
approximation.

Our motivation for this study is the desire to publicize progress made
in understanding the Gaussian approximation, which is often used in 
finite temperature/density applications, \cite{ch98,ab97} albeit often 
in incomplete form, 
and the hope that this work will ultimately lead to the clarification of 
the scalar meson spectroscopy, and in particular of the so-called 
$\sigma$ meson. 
For this reason, the present study of the GML model must be considered 
as a methodological work providing preparatory for a full-fledged, 
$N_{f} = 3$ calculation. 

This paper consists of six sections. In \S~2 we introduce the linear $\Sigma$ model. In \S~3
we outline the Gaussian approximation, and in \S~4 we demonstrate 
its chiral invariance. 
In \S~5 we present results of the numerical solution of the gap and the Bethe-Salpeter equations 
and analyze the solutions. 
Finally, we summarize and draw conclusions in \S~6. 

\section{The linear $\Sigma$ model}
\label{sec:linsm}

We confine ourselves to the $O(N=4)$ symmetric
linear $\sigma$ model for the sake of simplicity. 
The Lagrangian density of this theory is 
\begin{eqnarray}
{\cal L} =  \frac{1}{2} \left(\partial_{\mu} 
{\mbf \phi}\right)^{2} - V({\mbf \phi}^{2}) ~,
\label{e:lag}
\end{eqnarray}
where 
$${\mbf\phi} = (\phi_{0},\phi_{1},\phi_{2},\phi_{3})
= (\sigma, {\mbf \pi})$$ 
is a column vector and
$$V({\mbf \phi}^{2}) = - {1 \over 2} \mu_{0}^{2}
{\mbf \phi}^{2} + {\lambda_{0} \over 4}
\left({\mbf \phi}^{2}\right)^{2}. $$
We assume here that $\lambda_{0}$ and $\mu_{0}^{2}$ are not only 
positive, but such 
that spontaneous symmetry breakdown (SSB) occurs in the mean-field 
approximation [MFA], to be introduced below. 
As the chiral symmetry breaking ($\chi$SB) term in the Lagrangian, we choose 
\begin{equation}
{\cal L}_{\chi SB} = - {\cal H}_{\chi SB} = \varepsilon \sigma ,
\label{e:chisb}
\end{equation}
as suggested by the underlying NJL quark model. 
The interaction potential in the new field variable $s \equiv \sigma - \langle \sigma \rangle_{0 {\rm B}}$ reads
\begin{eqnarray}
V &=& 
{1 \over 2} \left(m_{\sigma {\rm B}}^{2} s^{2} + 
m_{\pi {\rm B}}^2 {\mbf \pi}^{2}\right) + 
\left({m_{\sigma {\rm B}}^{2} - m_{\pi {\rm B}}^2 
\over{2 f_{\pi {\rm B}}}}\right)
s \left(s^2 + {\mbf \pi}^{2}\right) \nonumber \\
& &+ 
\left({m_{\sigma {\rm B}}^{2} - m_{\pi {\rm B}}^2 \over{8 f_{\pi {\rm B}}^2}}
\right) 
\left(s^2 + {\mbf \pi}^{2}\right)^{2}. \
\label{e:potnc}
\end{eqnarray}
The scalar meson ($\sigma$) mass, the vacuum expectation value (v.e.v.)
and the pion (${\mbf \pi}$) mass are
\begin{subequations}
\begin{eqnarray}
\langle \sigma \rangle_{0 {\rm B}} &=& v_{\rm B} = f_{\pi {\rm B}} 
= - {\varepsilon \over \mu_{0}^2} 
+ \lambda_{0} {v_{\rm B}^{3} \over \mu_{0}^2}~,
\label{e:vacb}\\
m_{\sigma {\rm B}}^2 &=& - \mu_{0}^2 + 3 \lambda_0 f_{\pi {\rm B}}^2~, 
\label{e:sigb} \\
m_{\pi {\rm B}}^2 &=& - \mu_{0}^2 + \lambda_0 f_{\pi {\rm B}}^2 = 
{\varepsilon \over v_{\rm B}}~. 
\label{e:pib} \
\end{eqnarray}
\end{subequations}
Note that once the pion mass $m_{\pi {\rm B}}$ and decay constant 
$f_{\pi {\rm B}}$ have been fixed, there 
is only one free parameter left in this (tree) approximation, the scalar meson 
$\sigma$ mass $m_{\sigma {\rm B}}$. The quartic coupling constant $\lambda_{0}$
can be expressed as
\begin{eqnarray}
\lambda_{0} &=& \left({m_{\sigma {\rm B}}^{2} - m_{\pi {\rm B}}^2 
\over{2 f_{\pi {\rm B}}^{2}}} \right)~.
\label{e:vac1}\
\end{eqnarray}
This relation is a powerful result, as it implies a cubic dependence of the
$\sigma$ decay width on the its mass: 
\begin{eqnarray}
\Gamma_{\sigma\pi\pi} &=& 3 {\left(m_{\sigma {\rm B}}^{2} - m_{\pi {\rm B}}^2 
\right)^{2} \over{32 \pi f_{\pi {\rm B}}^{2} m_{\sigma {\rm B}}}}
\sqrt{1 - \left({2 m_{\pi {\rm B}}\over{m_{\sigma {\rm
B}}}}\right)^{2} }~.
\label{e:width}\
\end{eqnarray}
Therefore, as soon as the $\sigma$ mass exceeds 
the two-pion threshold, the decay width increases so quickly that instead 
of a sharp ``peak'' in the cross section there is a broad ``bump'', 
unrecognizable as a resonance. 
This fact is a consequence of the strong coupling implied
by Eq.~(\ref{e:vac1}), which in turn is a consequence of the linear 
realization of the chiral symmetry in the Born approximation.
Here, natural (and long-standing) questions arise: Does this effect survive 
after taking into account of the loop corrections? What kind of a 
particle, if any, corresponds to the $\sigma$ field, and how does one 
identify it? Many studies have 
been devoted to answering these questions, but most suffer from either being 
perturbative, which is unacceptable in the strong coupling 
case, or from not being chirally symmetric. We satisfy both of 
these requirements by employing the non-perturbative, chirally symmetric method 
described in the next section.

\section{The Gaussian variational method}

Over the past 20 years we have seen the relativistic Rayleigh-Ritz variational
approximation based on the Gaussian ground state (vacuum) functional 
elevated from a little-known specialist technical tool 
\cite{bg80} to a textbook method \cite{hat92}. 
This method sometimes also goes by the names    
'self-consistent mean field approximation' (MFA) and 'Hartree + RPA'.
\footnote{Due to the Bose statistics of our fields and
the covariance of our approach, RPA might be equivalent to
the Tamm-Dancoff 
approximation (TDA) in this case.}
In the following we use these terms interchangeably. 

\subsection{The basics of the Gaussian functional approximation}
\label{sec:gauss}

We use the Gaussian ground state functional Ansatz 
\begin{eqnarray}
\Psi_{0}[{\vec \phi}] = {\cal N} \exp \left( - {1 \over 4 \hbar} \int 
d{\bf x} \int d{\bf y} \left[\phi_{i}({\bf x}) - \langle  \phi_{i}
({\bf x}) \rangle \right] 
G_{ij}^{-1}({\bf x},{\bf y})\left[\phi_{j}({\bf y}) - \langle \phi_{j} 
({\bf y})) \rangle \right]\right)~,\nonumber\\
\label{e:gaus}
\end{eqnarray}
where ${\cal N}$ is the normalization constant, $\langle \phi_{i}({\bf x}) 
\rangle$ 
is the vacuum expectation value (v.e.v.) of the $i$-th spinless field (which 
henceforth we will assume to be translationally invariant, 
$\langle  \phi_{i}({\bf x}) \rangle = \langle  \phi_{i}(0) \rangle \equiv 
\langle  \phi_{i} \rangle$), and
$$G_{ij}({\bf x},{\bf y}) = {1 \over 2} \delta_{ij} \intop\limits {d {\bf k} 
\over (2 \pi)^{3}} {1 \over \sqrt{{\bf k}^{2} + m_{i}^{2}}} e^{i {\bf k} 
\cdot ({\bf x} - {\bf y})} .$$ 
Furthermore, note that we have explicitly kept $\hbar$ (while setting the 
velocity of light $c = 1$) to keep track of quantum corrections and count 
the number of ``loops" in our calculation. Then the ``vacuum'' (ground 
state) energy density becomes
\begin{eqnarray}
{\cal E}(m_{i}, \langle  \phi_{i} \rangle) &=& 
 - \varepsilon \langle \phi_{0} \rangle - {1 \over 2} \mu_{0}^{2} 
\langle {\mbf\phi} \rangle^{2} 
+ {\lambda_{0} \over 4}  
\left[\langle {\mbf\phi} \rangle^{2}\right]^{2}
\nonumber \\
&& +~ \hbar \sum_{i} \left[I_{1}(m_{i}) - {1 \over 2} \mu_{0}^{2}I_{0}(m_{i})
- {1 \over 2} m_{i}^{2} I_{0}(m_{i}) \right] 
\nonumber \\
&& +~
{\lambda_{0} \over 4} \Bigg\{ 
6 \hbar \sum_{i} \langle \phi_{i} \rangle^{2} I_{0}(m_{i}) 
+ 2 \hbar \sum_{i \neq j} \langle \phi_{i} \rangle^{2} I_{0}(m_{j}) 
\nonumber \\
&& +~  
3 \hbar^{2} \sum_{i} I_{0}^{2}(m_{i}) 
+ 2 \hbar^{2} \sum_{i < j} I_{0}(m_{i}) I_{0}(m_{j}) \Bigg\} \ ~, 
\label{e:ener}
\end{eqnarray}
where 
\begin{eqnarray}
I_{0}(m_{i}) &=& {1 \over 2} \int {d{\bf k} \over (2
\pi)^{3}} 
{1 \over \sqrt{
{\bf k}^{2} + m_{i}^{2}}} = 
i \int {d^{4} k \over (2 \pi)^{4}} 
{1 \over {k^{2} - m_{i}^{2} + i \epsilon}} =
G_{ii}({\bf x},{\bf x}) ~, 
\label{e:I_0} \\
 I_{1}(m_{i}) &=& {1 \over 2} \int {d {\bf k} \over (2
\pi)^{3}} 
\sqrt{{\bf k}^{2} + m_{i}^{2}} = 
- {i \over 2} \int {d^{4} k \over (2\pi)^{4}} 
\log \left(k^{2} - m_{i}^{2} + i \epsilon \right) +
{\rm const.}
\label{e:I_1} \
\end{eqnarray}
We identify $\hbar I_{1}(m_{i})$ with the 
familiar ``zero-point" energy density of a free scalar field of mass $m_{i}$. 

The divergent integrals $I_{0,1}(m_{i})$ are understood to be regularized 
via a UV momentum cutoff $\Lambda$. Thus we have introduced a new 
free parameter into the calculation. This was bound to happen in one 
form or another, since even in the renormalized perturbation theory one must 
introduce a new dimensional quantity (the ``renormalization scale/point'') 
at the one loop level. We treat this model as an effective theory and 
thus keep the cutoff without renormalization. \footnote{There are 
several renormalization schemes for the Gaussian approximation, but as
their names (``precarious'' and ``autonomous'') suggest, they are unstable.
\cite{stev84}}

\subsection{The gap equations}
\label{sec:gap1}

We vary the energy density with respect to the field vacuum expectation values 
$\langle \phi_{i} \rangle$ and the ``dressed" masses $m_{i}$. 
The extremization condition with respect to the field vacuum expectation 
values reads
$$\left({\partial {\cal E}(m_{i}, \langle  \phi_{i} \rangle) \over 
\partial 
\langle \phi_{i} \rangle}\right)_{min} = 0; ~~~i = 0,\ldots 3~ $$
or
\begin{eqnarray}
\left({\partial {\cal E}(m_{i}, \langle  \phi_{i} \rangle ) 
\over{\partial\langle  \phi_{0} \rangle}}\right)_{min} 
&=& 
- \varepsilon +
\langle  \phi_{0} \rangle \left[- \mu_{0}^{2} + {\lambda_{0}} \left(
\langle {\mbf \phi} \rangle^{2} 
+ 3 \hbar I_{0}(m_{0}) 
+ \hbar \sum_{i=1}^{3} I_{0}(m_{i}) \right) \right]_{min} = 0 
\nonumber \\
\left({\partial {\cal E}(m_{i}, \langle  \phi_{i} \rangle ) 
\over{\partial\langle \phi_{j=1,2,3} \rangle}}\right)_{min} 
&=& 
\langle  \phi_{j} \rangle \left[- \mu_{0}^{2} +  {\lambda_{0}} \left(
\langle {\mbf \phi} \rangle^{2}
+ \hbar  \sum_{j \neq k = 0}^{3}I_{0}(m_{k}) + 
3 \hbar I_{0}(m_{j}) \right) \right]_{min} = 0 ~.\nonumber \\
\label{e:vac2} 
\end{eqnarray}
Note that if we assume that $\langle \phi_{0} \rangle$ and 
$\langle \phi_{1,2,3} \rangle$ are simultaneously nonzero in the 
chiral limit $\varepsilon \rightarrow 0$, then after subtracting 
one of the equations (\ref{e:vac2}) from the other, we are forced to conclude 
that $I_{0}(M = m_{0}) = I_{0}(\mu = m_{1,2,3})$, or 
that these two masses are identical.
This, however, leads one to the symmetric 
phase of the theory, so we must ignore this possibility. Instead, we assume 
only one $\langle  \phi_{i} \rangle$ to be nonzero, e.g.  
$\langle \phi_{0} \rangle = v \neq 0$, while $\langle  \phi_{1,2,3} \rangle = 0$. 
Thus the first set of energy minimization equations in the Gaussian
variational approximation \cite{dms96} reads
\begin{subequations}
\begin{eqnarray}
\mu_{0}^{2} &=& - {\varepsilon \over v} +
{\lambda_{0}} \left[v^{2} +
3 \hbar I_{0}(M) + 3 \hbar I_{0}(\mu) \right] ~,
\label{e:veva} \\
\langle  \phi_{i} \rangle  &=& 0 ~~~i = 1,2,3~\ ~,
\label{e:vevb} 
\end{eqnarray}
\end{subequations}
where the divergent integral $I_{0}(m_{i})$ is understood to be 
regularized via a UV momentum cutoff $\Lambda$, either three dimensional or 
four dimensional. 

\begin{figure}
\epsfxsize = 10 cm   
\centerline{\epsfbox{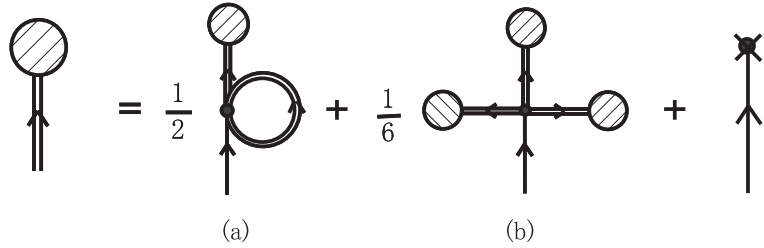}}
\caption{The one-point Green function Schwinger-Dyson equation determining the 
dynamics of the $\sigma$ model in the Hartree approximation:
the one-loop graph (a), and the tree tadpole diagram (b). 
The solid line denotes the bare meson multiplet, and the double solid line is 
the dressed meson multiplet. The shaded blob together with the double line 
leading to it (the ``tadpole") denotes the vacuum expectation value of the 
field (i.e. the one-point Green function), and the solid dot in the 
intersection of the four lines 
denotes the bare four-point coupling. The diagrams are explicitly multiplied 
by their symmetry numbers.}
\label{f:vev}
\end{figure}

\begin{figure}
\epsfxsize = 10 cm   
\centerline{\epsfbox{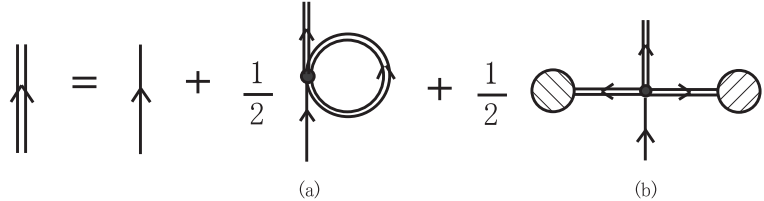}}
\caption{Two-point Green function Schwinger-Dyson equation: 
the one-loop graph (a), and the tree tadpole diagram (b). 
The symbols have the same meaning as in Fig. 1.}
\label{f:gap}
\end{figure}

Equations (\ref{e:veva}) and (\ref{e:vevb}) can be identified with the
truncated Schwinger-Dyson 
(SD) equations \cite{hat92} for the one-point Green function 
(see Fig. \ref{f:vev}).
We associate the nonvanishing vacuum expectation value (v.e.v.) 
with the ``sigma meson" field $\phi_{0}$, whose apparent mass 
is given by $m_{0} = M$, 
and the remaining three fields $\phi_{i} (i = 1,2,3)$, of mass 
$m_{i} = \mu$, form the pion triplet. 
The second set of energy minimization equations reads 
\begin{subequations}
\begin{eqnarray}
M^{2} 
&=& 
- \mu_{0}^{2} + {\lambda_{0}} 
\left[2 \langle \phi_{0} \rangle^{2} 
+ \langle {\mbf \phi} \rangle^{2} 
+ 3 \hbar I_{0}(M) + 3 \hbar I_{0} (\mu) \right] 
\nonumber \\
&=& 
- \mu_{0}^{2} + {\lambda_{0}} \left[3 v^{2} 
+ 3 \hbar I_{0}(M) + 3 \hbar I_{0} (\mu) \right] 
\label{e:gapa} \\
\mu^{2} &=& 
- \mu_{0}^{2} + 
{\lambda_{0}} \left[\langle {\mbf \phi} \rangle^{2}
+ \hbar I_{0}(M) + 5 \hbar I_{0}(\mu) \right] 
\nonumber \\
&=& - \mu_{0}^{2} + {\lambda_{0}} 
\left[v^{2} + \hbar I_{0}(M) + 5 \hbar I_{0}(\mu) \right] 
\label{e:gapb} \ ~.
\end{eqnarray}
\end{subequations}
Equations (\ref{e:gapa}) and (\ref{e:gapb}) also have the Feynman-diagrammatic 
interpretation shown in Fig. \ref{f:gap}.
Upon inserting Eqs. (\ref{e:veva}) and (\ref{e:vevb}) into
Eqs. (\ref{e:gapa}) and (\ref{e:gapb}),  
the following two coupled ``gap" equations emerge: 
\begin{subequations}
\begin{eqnarray}
M^{2} &=&  
{\varepsilon \over v} + 2 {\lambda_{0}} v^{2}  
\label{e:gap1a} \\
\mu^{2} &=&  
{\varepsilon \over v}  
+ 2 \lambda_{0} \hbar \left[I_{0}(\mu) - I_{0}(M) \right]  ~. \
\label{e:gap1b}
\end{eqnarray}
\end{subequations}
One might be tempted to identify $\mu$ with the pion mass and $M$ with 
the $\sigma$ mass, then solve these equations and stop there. However, 
with $\varepsilon = 0$
these equations admit only massive solutions $M > \mu > 0$ 
for real, positive values of $\lambda_{0}$ and $\mu_{0}^{2}$ and any  
real ultraviolet cutoff $\Lambda$ in the momentum integrals 
$I_{0}(m_{i})$ and $I_{1}(m_{i})$ as these are positive definite (for any real mass). 
In other words the ``pion" ($\phi_{1}, \phi_{2}$ and $\phi_{3}$)  
excitations are massive, with mass $\mu \neq 0$, in MFA, even 
in the chiral limit.
This looks like a breakdown of the $O(4)$ invariance of this method,
but, as discussed at length in Ref.~\citen{dms96}, 
there is a simple solution obtained using the Bethe-Salpeter equation.
\footnote{It is well known from the quantum 
many-body literature that the Hartree or mean-field approximation
does not respect internal symmetries. The corrective measure goes by 
the name of random phase approximation (RPA).} 
Before proceeding to solve the gap equations (\ref{e:gap1a}) and (\ref{e:gap1b}), we will have to 
determine the value of the $\varepsilon$ parameter in terms of observables
calculated in the Gaussian approximation. For this purpose we  
also have to use the Bethe-Salpeter equation.

\subsection{The Bethe-Salpeter equation (``RPA''): $\sigma \pi$ scattering}
\label{sec:bse1}

\begin{figure}
\epsfxsize = 10 cm   
\centerline{\epsfbox{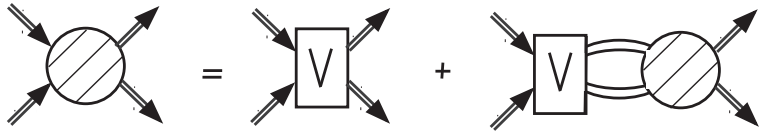}}
\caption{Four-point Green function Schwinger-Dyson or Bethe-Salpeter 
equation. The square ``box'' represents the potential, and the 
round ``blob'' is the BS amplitude itself. All lines represent 
dressed fields like the double lines in Figs. 1 and 2.} 
\label{f:bse}
\end{figure}

\begin{figure}
\epsfxsize = 10 cm  
\centerline{\epsfbox{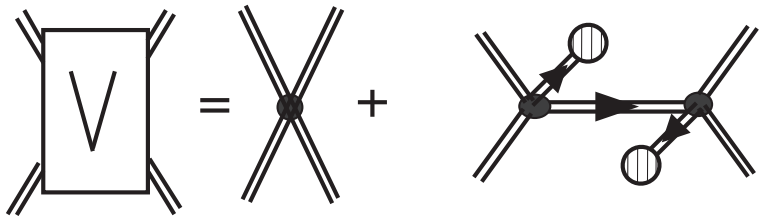}}
\caption{The potential (square ``box'') entering the Bethe-Salpeter 
equation, as defined in the RPA.} 
\label{f:pot}
\end{figure}

In Ref.~\citen{dms96} we have shown that the Nambu-Goldstone particles appear 
as poles in 
the {\it two-particle propagator} i.e. they are bound states of the two 
distinct massive elementary excitations in the theory. 
We specify the two-body dynamics in the theory in terms of the
four-point SD equation or, equivalently, of the Bethe-Salpeter equation, see 
Figs. \ref{f:bse},\ref{f:pot}.
We focus on the $s$-channel part $D_{\pi}(s)$ of the total 
four-point scattering amplitude $T(s,t,u)$. Its four-point SD equation reads 
\begin{eqnarray}
D_{\pi}(s) &=& V_{\pi}(s) + V_{\pi}(s) \Pi_{\pi}(s) D_{\pi}(s) ~,
\label{e:bsepi} \\
\Pi_{\pi}(s) 
&=& I_{M\mu}(s) = 
i \hbar \int {d^{4} k \over (2 \pi)^{4}} 
{1 \over {\left[k^{2} - M^{2} + i \epsilon \right]
\left[(k - P)^{2} - \mu^{2} + i \epsilon \right]}} ~,
\label{e:pols} \\
V_{\pi}(s) &=& 2 \lambda_{0} \left[1 + 
\left({2 \lambda_{0} v^{2} \over{s - \mu^{2}}}\right) \right] 
= 2 \lambda_{0} 
\left[1 + {M^{2} - {\varepsilon \over v} \over{s - \mu^{2}}}\right] ~,
\label{e:pot}\
\end{eqnarray}
with the solution
\begin{eqnarray}
D_{\pi}(s) = {V_{\pi}(s) \over{1 - V_{\pi}(s) \Pi_{\pi}(s)}} ~,
\label{e:sol}
\end{eqnarray}
where $s = (p_{1} + p_{2})^{2} \equiv P^{2}$ is the center-of-mass 
(CM) energy.
This propagator can also be written in the 
form 
(see Ref.~\citen{gross})
\begin{eqnarray}
D_{\pi}(s) &\simeq& 
{g_{\pi \phi_{i} \phi_{0}}^{2} \over{s -
m_{\pi}^{2}}}, 
\label{e:proppi}
\end{eqnarray}
where 
\begin{eqnarray}
g_{\pi \phi_{i} \phi_{0}}^{- 2} &\equiv& g_{eff}^{- 2}
= \left({1 \over{2 \lambda_{0}(M^{2} -
\mu^{2})}}\right)\Bigg\{
\left({M^{2} \over{M^{2} - \mu^{2}}}\right) 
\nonumber \\
&+& {1 \over{(4 \pi)^{2}}} \left[{1 \over 2} + 
\left({\mu^{2} \over{M^{2} - \mu^{2}}} \right)\left(1
- 
\left({M^{2} \over{M^{2} - \mu^{2}}}\right) 
\log \left({M^{2} \over
\mu^{2}}\right)\right)\right]\Bigg\} 
\nonumber \\
&\simeq&
\left({M^{2} \over{2 \lambda_{0} (M^{2} -
\mu^{2})^{2}}}\right)
= \left({v \over{M^{2} - \mu^{2}}}\right)^{2} .
\label{e:gpi}
\end{eqnarray}
We see that the second term in curly brackets is
roughly 1\%
as large as the first one, and therefore we may neglect it in the first
approximation.

Here we have simply presented the correct form of the four-point 
SD equation based on the truncation of the exact SD equation. 
 \cite{hat92} The derivation from the Gaussian 
approximation in the symmetric phase of the theory can be found in 
Ref.~\citen{bg80}. 
The corresponding derivation in the asymmetric (Nambu-Goldstone) 
phase can be found in Ref.~\citen{kerm98}. Furthermore, this BS 
equation is also the ``random phase approximation'' (RPA) equation of motion
that describes ``quasi-particles'' in this theory (see Ref.~\citen{ab97,kerm98,fw71}).

\subsection{The Bethe-Salpeter equation: $\pi \pi$ scattering}
\label{sec:bse2}

The dynamics are specified in terms of the SD, or equivalently, 
Bethe-Salpeter (BS) equation for the four-point Green functions
$D_{ij}(s)$, where the indices $i$ and $j$ denote the isospin of the pions 
in the initial and final states, respectively. 
We use the isospin invariance to split this
$9 \times 9$ matrix equation into three invariant subspaces:
(a) isoscalar, (b) isovector, and (c) isotensor. Because we consider $S$-wave 
scattering, the isovector amplitude vanishes identically, due to the 
Bose-Einstein statistics of the pions. The isotensor BS equation is linear 
and can be solved straightforwardly, but it is without distinguishing features. 
On the other hand, in the isoscalar channel we expect to see the 
$\sigma$ meson. The corresponding BS equations. consist of four
coupled equations that can be put into $2 \times 2$ matrix form. \cite{vd98}

The main difference between the isoscalar channel and the pion
channel, considered in \S~\ref{sec:bse1}, 
is that here we have two distinct intermediate states, one with two
``elementary'' sigma fields ($\phi_{0}$) and another with two 
``elementary'' pion fields ($\phi_{i}, i=1,2,3$). 
The isoscalar SD equations couple these two channels: 
\begin{eqnarray}
D_{MM}(s) &=&  V_{MM}(s) 
+ {1 \over 2} V_{MM}(s) I_{MM}(s) D_{MM}(s) 
+ {3 \over 2} V_{M\mu}(s) I_{\mu \mu}(s) D_{\mu M}(s) ~, \nonumber \\
D_{M\mu}(s) &=& V_{M\mu}(s) 
+ {1 \over 2} V_{M\mu}(s) I_{\mu \mu}(s) D_{\mu \mu}(s) 
+ {1 \over 2} V_{MM}(s) I_{MM}(s) D_{M\mu}(s) ~, \nonumber \\
D_{\mu M}(s) &=& V_{\mu M}(s) 
+ {1 \over 2} V_{\mu M}(s) I_{MM}(s) D_{MM}(s) 
+ {1 \over 2} V_{\mu \mu}(s) I_{\mu \mu}(s) D_{\mu M}(s) ~, \nonumber \\
D_{\mu \mu}(s) &=& V_{\mu \mu}(s) 
+ {1 \over 2} V_{\mu \mu}(s) I_{\mu \mu}(s) D_{\mu \mu}(s) 
+ {3 \over 2} V_{\mu M}(s) I_{MM}(s) D_{M\mu}(s)  ~. \
\label{e:bse}
\end{eqnarray}
The equations in (\ref{e:bse}) can be cast into matrix form
\begin{eqnarray} 
{\bf D}_{\sigma} = {\bf V} + \displaystyle
\frac{1}{2}{\bf V}~{\bf \Pi}~ 
{\bf D}_{\sigma} ~, 
\label{e:matrixbse}
\end{eqnarray} 
\begin{eqnarray} 
{\bf D}_{\sigma} &\equiv& 
\left(\begin{array}{cc} 
D_{MM} & D_{M\mu} \\ 
D_{\mu M} & {1 \over 3}\,D_{\mu\mu} 
\end{array}\right) ~,  
\label{e:matrixdef}
\end{eqnarray}
where the boldfaced symbols are matrices. 
The solution to the matrix equation
(\ref{e:matrixbse}) is 
\begin{eqnarray} 
{\bf D}_{\sigma} = (1 - \displaystyle \frac{1}{2}{\bf
V}~{\bf 
\Pi})^{-1}~{\bf V} ~,
\label{e:matrixsol}
\end{eqnarray} 
where 
\begin{eqnarray} 
\bf{V} &=& 
\left(\begin{array}{cc} 
V_{MM} & V_{M\mu}\\ 
V_{\mu M} & {1 \over 3}\,V_{\mu \mu} 
\end{array}\right)  
\nonumber  \\ 
&=& 
2 \lambda_{0} 
\left(\begin{array}{cc} 
3 \left[1 + 3 {M^{2} - {\varepsilon \over v} \over s - M^{2}} \right] & 
\left[1 + 3 {M^{2} - {\varepsilon \over v} \over s - M^{2}} \right] \\ 
& \\
\left[1 + 3 {M^{2} - {\varepsilon \over v} \over s - M^{2}} \right] & 
\left[5 + 3 {M^{2} - {\varepsilon \over v} \over s - M^{2}} \right] \\ 
\end{array}\right) ~,
\label{e:potmatrix}
\end{eqnarray} 
and 
\begin{eqnarray} 
{\bf \Pi} &=& 
\left(\begin{array}{cc} 
I_{MM} & 0 \\ 
0 & 3\,I_{\mu \mu} 
\end{array}\right) ~.  
\label{e:imatrix}
\end{eqnarray}
The invariant functions $I_{ii}(s), J_{ii}(s)$ are
given by
\begin{eqnarray} 
I_{ii} &=&
i \int {d^{4} k \over (2 \pi)^{4}} 
{1 \over {\left[k^{2} - m_{i}^{2} + i \varepsilon
\right]
\left[(k - P)^{2} - m_{i}^{2} + i \varepsilon
\right]}}
\nonumber \\ 
&=& I_{ii}(0) + 
{s \over{(4 \pi)^2}} \left(-1 + J_{ii}(s)\right) ~, \
\label{e:ifns}
\end{eqnarray} 
where $s = P^2$, and the real and imaginary parts of
$J_{ii}(s)$ are
\begin{equation}
J_{ii}(s) = \left\{ \begin{array}{ll}
\displaystyle \sqrt{{4m^2_{i} \over s} - 1}\,
\arcsin \left(\sqrt{{s \over 4m_i^2}}\right)~,&
\displaystyle(0 <  {s \over 4m_i^2} < 1) \\
\displaystyle \sqrt{1 - {4m^2_{i} \over s}}\,
\left[\log\left(\sqrt{s \over 4m^2_{i}} 
+ \sqrt{{s \over 4m^2_{i}} - 1} \right) - i\,
\displaystyle {\pi \over 2}\right]~.&
\displaystyle (1 \leq {s \over 4m_i^2} < \infty ) 
\end{array}
\right.
\end{equation}

The equal mass integrals at zero momentum $I_{ii}(0)$ are cutoff dependent 
constants (``subtraction constants'' in the dispersion relations
language) whose values are determined by the solutions $M, \mu,$ and $\Lambda$
to the gap equation (\ref{e:gap2b}). Thus, fixing of the subtraction 
constants is one of the primary consequences of the self-consistent 
gap equations. We use
\begin{subequations}
\begin{eqnarray}
I^{\rm (4)}_{ii} (0) &=& (4 \pi)^{-2}
\left[{x_{4} \over{x_{4} + 1}} - \log(x_{4} + 1)\right]~; ~~
~ x_{4} = \left({\Lambda_{4} \over m}\right)^{2} ~,
\\
I^{\rm (3)}_{ii} (0) &=& 2 (4 \pi)^{-2}
\left[ {x_{3} \over{\sqrt{x_{3}(1 + x_{3})}}} - 
{\rm ln}(\sqrt{x_{3}} + \sqrt{1 + x_{3}})\right];~
x_{3} = \left({\Lambda_{3} \over m}\right)^{2}
\ .
\end{eqnarray}
\end{subequations}
Thus self-consistency of the gap equations enters 
into the scattering problem and constrains the remaining free parameters. 

We see from Eq. (\ref{e:bse}) that in order to calculate the $D_{MM}$ 
amplitude, we need to know the $D_{\mu M}(s)$ amplitude, and {\it vice versa}. 
In other words, we must solve the system of four coupled equations (\ref{e:bse}). 
It turns out that this system splits into two systems with two
unknowns, with the same discriminant $\cal D$. This fact, 
(a) ensures that there are at most two poles in the solutions, and 
(b) greatly simplifies the algebra. 
The solutions to the equations (\ref{e:bse}) are 
\begin{eqnarray}
D_{MM}(s) &=& 
{1 \over{{\cal D}(s)}}
\left(V_{MM}(s) - 12 \lambda_{0} I_{\mu \mu}(s) V_{M\mu}(s)
\right)~,
\nonumber \\
D_{\mu \mu}(s) &=& 
{1 \over{{\cal D}(s)}}
\left(V_{\mu \mu}(s) - 12 \lambda_{0} I_{MM}(s) V_{M\mu}(s)
\right)~,
\nonumber \\
D_{\mu M}(s) &=& D_{M\mu}(s) = 
{1 \over{{\cal D}(s)}} V_{M\mu}(s)
= {1 \over{{\cal D}(s)}} V_{\mu M}(s) ~,
\label{e:bsesol}
\end{eqnarray}
where 
\begin{eqnarray}
{\cal D}(s) &=&
1 - {1 \over 2} 
\left[V_{MM}(s) I_{MM}(s) + V_{\mu \mu}(s) I_{\mu \mu}(s) \right]
+ 6 \lambda_{0} I_{MM}(s) I_{\mu \mu}(s) V_{\mu M}(s) 
\label{e:d}
\end{eqnarray} 
is the discriminant of ``one half" of the system of equations (\ref{e:bse}). 
Elementary and composite states in the $s$-channel 
manifest themselves as poles in the ${\bf D}_{\sigma}(s)$ matrix, 
or equivalently as roots of 
\begin{eqnarray}
(s - M^2){\cal D}(s) &=& 0. 
\label{e:mass}
\end{eqnarray} 
This equation is identical to the formula for the $\sigma$ mass in the 
optimized perturbation theory (OPT)~\cite{okop96} 

An inspection of Eq. (\ref{e:d}) leads one to think that 
$(s - M^2){\cal D}(s)$ should be a quadratic polynomial (function) of 
$s$ and therefore have two roots, since each 
$(s - M^2) V_{ij}(s)$ is a linear function of $s$.
However, as a consequence of chiral symmetry, 
one finds that
the function $(s - M^2){\cal D}(s)$ is (only) linear in $s$. This, 
in turn, ensures that there is only one root of the ``eigenvalue" 
equation (\ref{e:mass}),
and hence only one pole in the $\sigma$ propagator at least for 
small values of $s$, where the $I_{ii}$ functions' logarithmic $s$ 
dependence may be neglected. 
For large values of $s$, this is no longer the case, as we show 
in \S~\ref{sec:num}.

\section{Chiral symmetry Ward identities in the Gaussian approximation}
\label{sec:chiWT}

Chiral Ward-Takahashi identities follow from the underlying chiral 
symmetry of the linear sigma model and typically relate (n-1)-point Green 
functions to other n-point functions and/or currents.
These identities were developed by Lee 
at the perturbative one loop level, \cite{lee69} and by Symanzik at 
arbitrary orders of perturbation theory. \cite{sym70} 
We call them the Lee-Symanzik [LS] identities. To our knowledge,
for selected infinite classes of diagrams in the linear sigma model 
no proofs of LS identities were given prior to Ref.~\citen{dms96}. 
The NG theorem is the simplest LS identity. When the chiral symmetry 
is explicitly broken, NG theorem turns into a relation between the 
chiral symmetry-breaking parameter and the NG boson 
mass, as first discussed by Dashen. \cite{dash69}
The NG theorem in the chiral limit has already been addressed in the 
Gaussian approximation and related formalisms in
Refs.~\citen{dms96,ab97} and \citen{tsue00}.
For this reason, we consider the nonchiral case. 

\subsection{Dashen's formula and the Nambu-Goldstone theorem}
\label{sec:ng}

As shown in Ref.~\citen{dms96} in the chiral limit the Nambu-Goldstone 
particle appears as a zero-mass pole in the pion channel 
{\it two-particle propagator} Eq.~(\ref{e:sol}).
Next, we consider the zero CM energy $P = 0$ polarization function 
$V_{\pi}(0) \Pi_{\pi}(0) $
in the nonchiral case. We use Eq. (\ref{e:gapb}) to write
\begin{eqnarray}
V_{\pi}(0) \Pi_{\pi}(0) 
&=& 
{2 {\lambda_{0}} \hbar \over (M^{2} - \mu^{2})} 
\left[I_{0}(M) - I_{0}(\mu)\right] \left[1 - 
{M^{2} - {\varepsilon \over v} \over \mu^{2}}
\right] \nonumber \\  
&=& 
\left({{\varepsilon \over v} - \mu^{2} \over{M^{2} - \mu^{2}}}\right) 
\left[1 - {M^{2} - {\varepsilon \over v} \over \mu^{2}} \right] \ ~
\label{e:polz2}
\end{eqnarray}
and then use Eq. (\ref{e:gapa}) to obtain the final result,
\begin{eqnarray}
V_{\pi}(0) \Pi_{\pi}(0) &=& 
1 - {\varepsilon \over v}
\left({M^{2} \over{\mu^{2}(M^{2} - \mu^{2})}}\right)
+ {\cal O}(\varepsilon^{2}) \ ~.
\label{e:polz3}
\end{eqnarray}
The propagator Eq. (\ref{e:sol}) evaluated at zero
momentum 
can be written as
\begin{eqnarray}
D_{\pi}(0) &\simeq& 
{g_{\pi \phi_{i} \phi_{0}}^{2} \over{- m_{\pi}^{2}}} 
= - \left({2 \lambda_{0} \over{m_{\pi}^{2}}}\right) 
\left({M^{2} - \mu^{2} \over{M^{2}}}\right)^{2}  
\nonumber \\
&=& {V_{\pi}(0) \over{1 - V_{\pi}(0) \Pi_{\pi}(0)}}
= - 2 \lambda_{0} \left({v \over{\varepsilon}}\right) 
\left({M^{2} - \mu^{2} \over{M^{2}}}\right)^{2} ,
\label{e:pimass}
\end{eqnarray}
which leads to the result $\varepsilon = m_\pi^2 v + 
{\cal O}(\varepsilon^{2})$.
This is also in agreement with a general result due to Dashen. \cite{dash69}
The vacuum expectation value of the $\Sigma$ operator, 
when sandwiched between two vacuum states, yields Dashen's formula, 
\begin{equation} 
\left(f m^{2} f \right)^{ab} = f_{a} m^{2}_{ab} f_{b} 
= - \langle 0 \left| 
\big[Q^{a}_{5}, [Q^{b}_{5}, {\cal H}_{\chi SB}]
\big] \right| 0 \rangle + {\cal O}(m_{\pi}^{4}) ~, \
\label{e:dash1}
\end{equation}
for the pseudoscalar meson mass squared ($m_{ps}^2$) and decay constant 
$f_{a}$
for an arbitrary chiral symmetry-breaking term 
in the  Hamiltonian density ${\cal H}_{\chi SB}$. 
This is a model-independent result based on the 
equations of motion in the Heisenberg representation
and on the LSZ reduction formulas. 
We  use the canonical commutation relations and the axial charge 
to evaluate the $\Sigma$ operator in the linear $\Sigma$ model, 
\begin{eqnarray}
\Sigma \delta^{ab} &=& 
\big[Q^{a}_{5}, [Q^{b}_{5}, {\cal H}_{\chi SB}(0)]\big] =
- \varepsilon \sigma \delta^{ab} ~, \
\label{e:dash3}
\end{eqnarray}
which vanishes when the chiral symmetry is not explicitly broken.
Taking the vacuum expectation value of this expression, we find
\begin{equation} 
\left(m_{\pi} f_{\pi} \right)^{2} = 
\varepsilon \langle 0 \left| \sigma \right| 0 \rangle 
+ {\cal O}(\varepsilon^{2}) ~,
\label{e:dash4}
\end{equation}
which leads to 
\begin{equation}
\varepsilon = m_{\pi}^{2} f_{\pi}~~.
\end{equation}
This relation is satisfied both in the Born and Gaussian 
approximations, which indicates that both are chirally symmetric.
Now that $\varepsilon$ has been fixed, note that the gap equation (\ref{e:gap1b}) 
implies $m_{\pi}^{2} \leq \mu^{2}$, in agreement with the variational nature 
of the Gaussian approximation. This is important in the 
numerical calculations discussed below and for the physical interpretation.

\subsection{Axial current Ward identity}
\label{sec:axwti}
\begin{figure}
\epsfxsize = 15 cm   
\centerline{\epsfbox{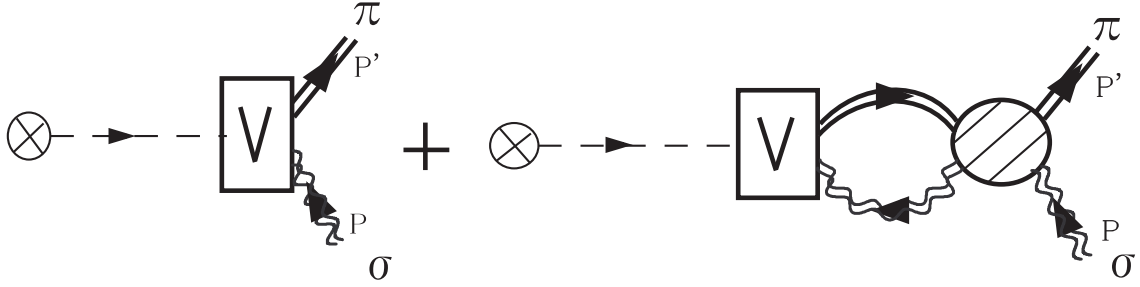}}
\caption{Axial current matrix element}
\label{e:axialcurrent}
\end{figure}
The axial current matrix element corresponding to the
Feynman diagrams
displayed in Fig.~\ref{e:axialcurrent} reads 
\begin{eqnarray}
J_{\mu 5}^{a} (p^{'}, p) &=& 
\langle \phi^{a} (p')|J_\mu(0)|\phi_{0}(p) \rangle 
\nonumber \\
&=& 
(p^{'} + p)_{\mu} + q_{\mu}
\left({M^{2} - {\varepsilon \over v} \over q^{2} -
\mu^{2}}\right)  
\nonumber \\
&-& 
\Gamma_{\mu 5} (q) D_{\pi}(q) ~, 
\label{e:ax1} \
\end{eqnarray}
where $\Gamma_{\mu 5} (q)$ is defined by 
\begin{eqnarray}
\Gamma_{\mu 5} (q) &=& 
i \int {d^{4} k \over (2 \pi)^{4}} 
\left[(2 k + q)_{\mu} + q_{\mu}
\left({M^{2} - {\varepsilon \over v} \over q^{2} - \mu^{2}}\right) 
\right]
{1 \over {\left[k^{2} - M^{2} \right]
\left[(k + q)^{2} - \mu^{2} \right]}} \
\label{e:ax}
\end{eqnarray}
and satisfies the chiral Ward identity, \cite{dm96} 
\begin{eqnarray}
q^{\mu} \Gamma_{\mu 5} (q) &=& 
\left[{\mu^{2} \over{2 \lambda_{0}}}
\left(V_{\pi}(0) \Pi_{\pi}(0) - V_{\pi}(q^{2}) \Pi_{\pi}(q^{2}) \right) 
- {\varepsilon \over v} \left(\Pi_{\pi}(0) - \Pi_{\pi}(q^{2}) \right) \right]
\ ~.
\label{e:wi1}
\end{eqnarray}
We insert the vertex $\Gamma_{\mu 5} (q)$ in Eq. (\ref{e:ax}) together with the 
two-body propagator $D_{\pi}(q^{2})$ in Eq. (\ref{e:sol}) into Eq. (\ref{e:ax1}) 
to find
\begin{eqnarray}
J_{\mu 5} (p^{'}, p) &=& 
(p^{'} + p)_{\mu} + q_{\mu}
\left({M^{2} - {\varepsilon \over v} \over q^{2} - \mu^{2}}\right)  
\nonumber \\
&&-~ 
{q_{\mu} \over q^{2}} \left[{\mu^{2} \over{2 \lambda_{0}}}
\left(V_{\pi}(0) \Pi_{\pi}(0) - V_{\pi}(q^{2}) \Pi_{\pi}(q^{2}) \right) 
- {\varepsilon \over v} \left(\Pi_{\pi}(0) - \Pi_{\pi}(q^{2}) \right) \right]
\nonumber \\
&&\times~ 
\left({V_{\pi}(q^{2}) \over{1 - V_{\pi}(q^{2}) \Pi_{\pi}(q^{2})}} \right) 
\nonumber \\
&&\simeq~ 
(p^{'} + p)_{\mu} + q_{\mu}
\left({M^{2} - \mu^{2} - m_{\pi}^{2} \over{q^{2} - m_{\pi}^{2}}}\right) ~. 
\label{e:ax3} \
\end{eqnarray}
Here we have $q^{\nu} = (p^{'} - p)^{\nu}$. 
This result manifestly lacks a pole at $q^{2} = \mu^2$.
The composite state plays precisely the role of the Nambu-Goldstone 
boson in the conservation of the (axial) Noether current, i.e., in the
basic axial Ward-Takahashi identity
\begin{eqnarray}
q^{\nu} J_{\nu 5} (p^{'}, p)  &=& 
\left(p^{'2} - \mu^{2}\right) -
\left(p^{2} - M^{2}\right) + m_{\pi}^{2}
\left({M^{2}- \mu^{2} - m_{\pi}^{2} \over q^{2} - 
m_{\pi}^{2}} - 1 \right)~,
\label{e:elwi} \
\end{eqnarray}
which follows directly from Eq.~(\ref{e:ax3}).
Furthermore, $f_{\pi}$ is defined by 
\begin{eqnarray}
\langle 0 | J_{\mu 5} | \Pi({\bf q}) \rangle 
= f_{\pi}(q) q_{\mu} = g^{eff} \Gamma_{\mu 5}(q), 
\label{e:fpi} \
\end{eqnarray}
from which (in the chiral limit) it follows that
\begin{eqnarray}
f_{\pi}(0) g^{eff} = M^{2} - \mu^{2} ,
\label{e:fpi1} \
\end{eqnarray}
by way of the axial Ward identity Eq. (\ref{e:wi1}).
This result, together with Eq. (\ref{e:gpi}) for
$g^{eff}$ forms
the basic result for the composite pion decay constant,
\begin{eqnarray}
f_{\pi} = f_{\pi}(0) = g_{eff}^{-1} \left(M^{2} -
\mu^{2}\right)
= v \left[1 + {\cal O}\left((4 \pi)^{- 2}\right)
\right] \simeq v .
\label{e:fpi1} \
\end{eqnarray}

\section{Numerical solutions}
\label{sec:num}

\subsection{The self-consistency or gap equation}
\label{sec:gap}

Having determined the value of the parameter $\varepsilon = v m_{\pi}^{2}$ in 
terms of
observables, we can solve the gap equations. We fix the v.e.v. $v$ at
a value of $93$ MeV for the
pion decay constant $f_{\pi}$ and a value of $140$ MeV for the
physical pion mass $m_{\pi}$.
As a result, the system of gap equations (\ref{e:gap1a}) and (\ref{e:gap1b}) turns into a single equation:
\begin{eqnarray}
v^{2} &=& f_{\pi}^{2} = 
\left({M^{2} - {\varepsilon \over v} \over{\mu^{2} - {\varepsilon \over v}}}
\right)
\hbar \left[I_{0}(\mu) - I_{0}(M) \right]  
\nonumber \\
&=&  
\left({M^{2} - m_{\pi}^{2} \over{\mu^{2} - m_{\pi}^{2}}}\right)
\hbar \left[I_{0}(\mu) - I_{0}(M) \right]  
~. \
\label{e:gap2b}
\end{eqnarray}
Here, we have used Eq.~(\ref{e:I_0})
as the integral to be regulated.
We give here the results for (covariant) four-dimensional
Euclidean cutoff regularization 
the three-dimensional regularization of this quadratically
divergent integral:
\begin{subequations}
\begin{eqnarray}
I_{0}^{\rm (4)}(m^2) &=&
(4 \pi)^{-2}   m^2 \left[ x_{4} - {\rm ln}(1 + x_{4}) \right]~; ~~
~ x_{4} = \left({\Lambda_{4} \over m}\right)^{2} ~,
\\
I_{0}^{\rm (3)}(m^2) &=& 
2 (4 \pi)^{-2} m^2 
\left[\sqrt{x_{3}(1 + x_{3})} - 
{\rm ln}(\sqrt{x_{3}} + \sqrt{1 + x_{3}})\right];~ 
x_{3} = \left({\Lambda_{3} \over m}\right)^{2} \ .
\end{eqnarray}
\end{subequations}
Note that the $\hbar \to 0$ limit of Eq.~(\ref{e:gap2b}) 
is non-trivial: on the right-hand side both the numerator 
and the denominator vanish. 
Further, note that one parameter, in this case the cutoff $\Lambda$, 
remains free. 

\begin{figure}
\epsfxsize = 8 cm   
\centerline{\epsfbox{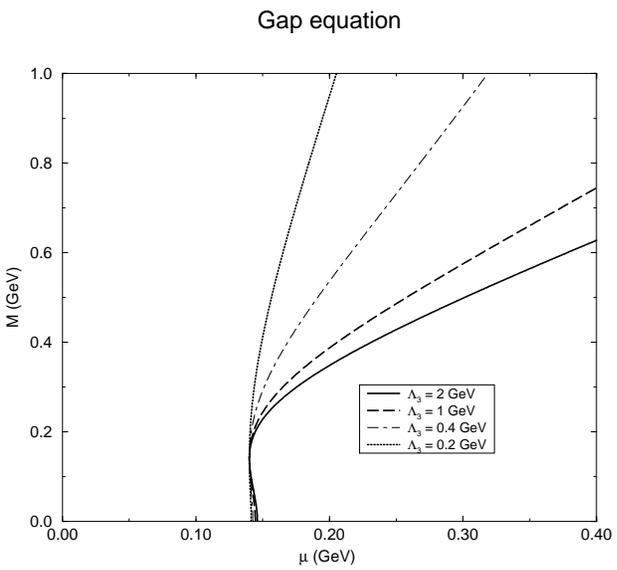}}
\caption{Solutions to the nonchiral gap equation
in the Gaussian approximation to the $O(4)$ 
linear sigma model with different values of the (three-dimensional) 
cutoff $\Lambda_{3}$.}
\label{f:gap1}
\end{figure}

\begin{figure}
\epsfxsize = 8 cm 
\centerline{\epsfbox{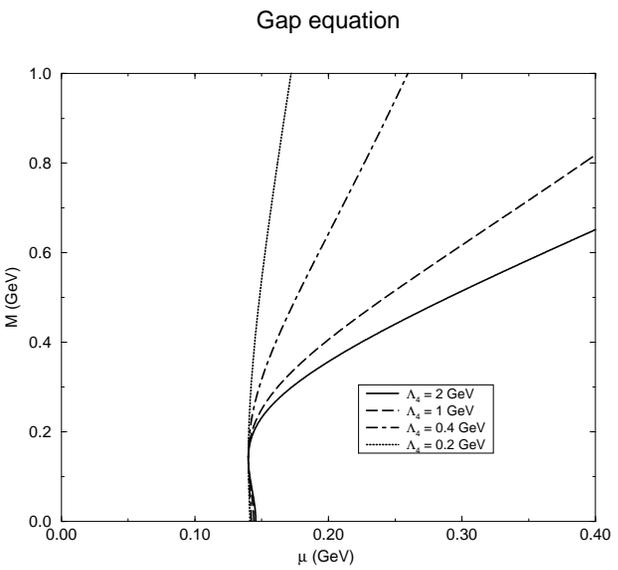}}
\caption{Solutions to the nonchiral gap equation
in the Gaussian approximation to the $O(4)$ 
linear sigma model with different values of the (four-dimensional) 
cutoff $\Lambda_{4}$.}
\label{f:gap2}
\end{figure}

Numerical solutions to the equation (\ref{e:gap2b}) are plotted in 
Figs. \ref{f:gap1} and \ref{f:gap2} for various 
values of the 
cutoff $\Lambda$.
Every point on the $(\mu, M)$ curve represents a solution to the gap 
equation, thus signalling the existence of 
freedom of choice in the form of one continuous free  
parameter. This free parameter can be related to the bare 
coupling constant $\lambda_{0}$ by
Eq. (\ref{e:gap1a}) for every $(\mu, M)$ pair. 

In Figs. \ref{f:gap1} and \ref{f:gap2} we see that as the cutoff 
$\Lambda$ increases, 
all solutions to the gap equation approach the symmetry restoration limit
$M \to \mu$ for large values of $M$, or equivalently large values of $\lambda_{0}$.
This implies that the large boson loop effects lead to symmetry restoration,
in contrast to the fermion loops, which lead to symmetry breaking.
Solutions that lie above the $2 \mu$ threshold require rather small 
values of (either kind of) cutoff $\Lambda$. However,
this is in agreement with the small (second) meson-loop cutoff found in $1/N_{c}$ 
studies of the NJL chiral quark model. \cite{onc95} This does not mean 
that the loop effects are necessarily small, however. 

\subsection{The Bethe-Salpeter or scattering Equation}
\label{sec:phase}

We reduced the BS equation (\ref{e:bse}) in the isoscalar channel to solving
a single algebraic equation (\ref{e:mass}) involving the transcendental 
analytic functions $I_{MM}(s)$ and $I_{\mu \mu}(s)$  
with branch cuts at imaginary parts above the corresponding thresholds. 
\footnote{$I_{MM}(s)$ and  $I_{\mu \mu}(s)$ have 
logarithmic branch points and therefore infinitely many sheets, in 
contrast with the nonrelativistic case, in which the branch points are of 
the square root type, with only two sheets.} 
This equation has, in general, real and imaginary parts: For $\sigma$
mass values lying below the two-body threshold, only the real part is 
relevant, while for heavier $\sigma$ masses the imaginary 
part must be taken into account as well. The latter determines the natural decay 
width of the $\sigma$ meson.

\begin{figure}
\epsfxsize = 8 cm 
\centerline{\epsfbox{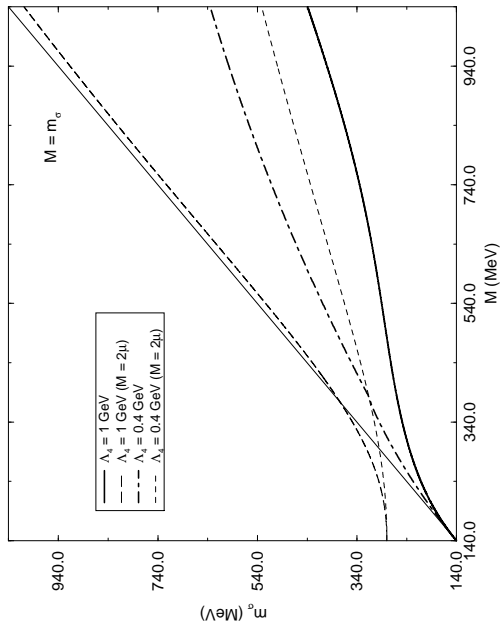}} 
\caption{The solution $m_{\sigma}$ to the Bethe-Salpeter equation in the 
isoscalar channel as a function of the variational parameter $M$ for 
various values of the cutoff, $\Lambda$. The curves denoted $2 \mu$
show the movement of the $2 \mu$ threshold for corresponding
values of the parameters $M$ and $\Lambda$.} 
\label{f:bsesol}
\end{figure}

From numerical solutions to the real part of Eq. (\ref{e:mass}) 
shown in Fig. \ref{f:bsesol},
it can be seen that the $\sigma$ mass is always shifted {\it downward} 
from the ``elementary" sigma field ($\phi_{0}$) mass $M$, in agreement 
with the variational property of the mean-field approximation. 
For small values of $M$ $(\leq 2 \mu)$ (i.e., 
when the coupling constant $\lambda_{0}$ 
lies below some critical value $\lambda_{c}$, 
which is a function of the masses $\mu, M$ and the cutoff $\Lambda$), 
the $\sigma$ meson mass (i.e., the real part of the pole 
position) drops below the $2 \mu$ threshold, and the $\sigma$ 
meson comes to consist predominantly of the bare $\phi_{0}$ state with some 
$2 \pi$ and $2 \sigma$ ``cloud'' components admixed to its wave function. 
\footnote{Another interpretation of these results (that might be only 
semantically 
different from that one) has been given in the language of operator many-body
(``quasi-particle RPA'') methods. \cite{ab97,fw71}}
With increasing coupling constant $\lambda_{0}$, the physical
$\sigma$'s mass increases above the $2 \mu$ threshold, 
and the ``bare'' and ``dressed'' components of the wave function 
can no longer be separated. Then, the
state itself must be considered as predominantly a meson-meson composite. 
For weak couplings, only one state has been found to exist in the 
$\sigma$ channel of the MFA.
In Fig. \ref{f:bsesol} it can be seen that the $\sigma$ mass 
changes continuously with decreasing coupling $\lambda_{0}$ 
and connects smoothly to the perturbative $\sigma$ mass in the weak 
coupling limit.

\begin{figure}
\epsfxsize = 8 cm 
\centerline{\epsfbox{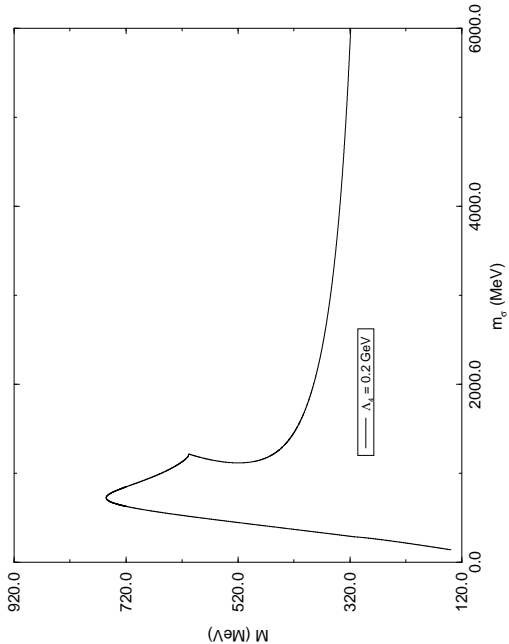}} 
\caption{The solution $m_{\sigma}$ to the Bethe-Salpeter equation in the 
isoscalar channel as a function of the variational parameter $M$ for 
a small cutoff, $\Lambda$ = 0.2 GeV. Note the 
double-valued nature of this functions, as well as the absence of
solutions above certain values of $M$.}
\label{f:bsesol2}
\end{figure}

Note, further, that for many values of the cutoff $\Lambda$ and 
above some critical value of $M$, far into the $2 \mu$
continuum, there is a second root of the real part of Eq. (\ref{e:mass}). 
As increases $M \sim \sqrt{\lambda_{0}}$, the two roots 
sometimes merge into one and then immediately disappear (see Fig \ref{f:bsesol2}). 
For other values of the parameters, the two roots diverge, the smaller one 
moving down to zero, while the heavier one moves back up in mass again. 
In either case, the smaller zero, which is connected to the perturbative
solution, has an upper limit generally below 1 GeV. This is 
perhaps the most interesting result of this paper. We remind the 
reader that these results are not artifacts of the Gaussian approximation 
viewpoint, as exactly the same equations govern the OPT meson masses.

The question of the physical meaning of the two zeros arises; i.e., if actual 
poles exist on the second unphysical Riemann sheet of the $S$-matrix, 
that can be associated with these zeros in the real part of 
the inverse propagator? The larger zero is almost certainly not a
conventional pole, because the derivative of Eq. (\ref{e:mass}) 
evaluated at the root
has sign opposite to that at the smaller root. 
Because this derivative is related to 
the effective coupling constant squared, this implies that the upper
state has a non-Hermitian coupling to the bare states. 
This question is more difficult to address, as it demands analytic
continuation onto the second Riemann sheet. It will be left to a 
future investigation.

\section{Summary and conclusions}
 
In summary, we have:
1) constructed a unitary, Lorentz and chirally invariant, self-consistent 
variational approximation to the linear $\sigma$ model;
2) solved the coupled self-consistent equations of motion in this, the
mean-field plus random-phase approximation (MFA + RPA). The solutions 
to these equations also determine the minimum of the ``optimized 
perturbation theory'' effective potential.;
3) shown that the particle content of the mean-field 
plus random-phase approximation
to the $O(4)$ linear sigma model is the same as in the Born
approximation, at least for weak coupling, i.e., 
there are three Goldstone bosons $\pi$ and one $\sigma$ state.
4) found that the pions' mass is unchanged, as a consequence of 
the validity of Dashen's relation in the MFA + RPA, whereas the
$\sigma$ mass and width can differ substantially from the Born values, depending on the free parameters.
5) calculated the $\sigma$ meson mass by solving 
the nonperturbative Bethe-Salpeter equation. We found a second
solution for large values of mass and coupling, whose physical 
interpretation is yet unclear. 

The mean-field or Gaussian method was initially fraught with problems 
when applied to 
the linear $\sigma$ model with spontaneously broken internal symmetry - 
the Goldstone theorem did not seem to ``work.''
This problem was solved in Ref. \citen{dms96}: The Goldstone boson found 
in the Gaussian approximation 
\cite{dms96} turns out to be a composite massless state, just as in 
the NJL model. Yet, there seemed to exist another massive 
state with the quantum numbers of the pion. 
In fact, however, this is true only in appearance: There is no pole
in the propagator corresponding to this ``particle''. 
The MFA to the bosonic linear $\sigma$ model is significantly different 
from the NJL one in one regard: Whereas in the NJL model,
the gap equation describes ``dressing'' of the fermions and the BS equation 
describes mesons as bound states of dressed fermions, in the linear $\sigma$ 
model both the gap and the BS equations describe (two different) ``dressings'' 
(first and second renormalizations) of mesons. 
Therefore, the results of the intermediate (first) renormalization remain in 
the theory even after the second renormalization and confuse the issues of 
the physical content. 
A more complicated situation exists in the scalar sector. 
We have not yet written the last word on this subject. 

We hope to extend these calculations to physical applications for 
more realistic Lagrangians in the future.

\section*{Acknowledgements}

One of the authors [V.D.] would like to acknowledge a Center-of-Excellence 
(COE) Professorship for the year 2000/1 and the hospitality of RCNP. We
wish to acknowledge the kind help in all matters concerning computers and software 
that we received from Ms. Miho Takayama-Koma, as well as help 
with Canvas software from Dr. F. Araki. The authors wish to thank
Professor H. Toki for introducing them.

\end{document}